\documentclass[preprint2]{aastex}
\usepackage{spr-astr-addons}
\usepackage{graphicx}
\usepackage{checkend}

\begin{document}
%Insert your title here
\title{Herbig-Haro Objects - Tracers of the Formation of Low-mass Stars and Sub-stellar Objects}

\shorttitle{HHOs from low-mass objects}        % if too long for running head
\shortauthors{Stecklum et al.}

\author{Bringfried Stecklum and Helmut Meusinger}
\affil{Th\"uringer Landessternwarte Tautenburg, Sternwarte 5, Tautenburg, 07778 Tautenburg, Germany}
\email{stecklum@tls-tautenburg.de}

\author{Dirk Froebrich}
\affil{Centre for Astrophysics \& Planetary Science, University of Kent, Canterbury CT2 7NH, U.K.}

\begin{abstract}
Herbig-Haro objects (HHOs) are caused by outflows from young objects. Since the outflow relies on mass accretion from a circumstellar disk, it indicates ongoing growth. Recent results of infrared observations yielded evidence for disks around brown dwarfs. {This suggests that at least a certain fraction of brown dwarfs forms like stars.} Thus, young sub-stellar objects might cause HHOs as well. We present selected results of a general survey for HHOs based on DSS-II plates and CCD images taken with the Tautenburg Schmidt telescope. Numerous young objects could be identified due to their association with newly detected HHOs. In some cases the luminosity is consistent with very low-mass stars or close to sub-stellar values. This holds for L1415-IRS and a few infrared sources embedded in other dark clouds (e.g., GF9, BHR111). The question on the minimum mass for outflow activity is addressed. 
\end{abstract}

\keywords{ISM: clouds, individual(GF9, LDN1415, BHR111), Herbig-Haro objects, individual(HH892), jets and outflows --- stars: formation, low-mass, brown dwarfs, individual(BHR111-IR, IRAS04376+5413, \newline IRAS20503+6006)}

\section{Introduction}
\label{intro}
Herbig-Haro objects (\cite{1950ApJ...111...11H}, \cite{1952ApJ...115..572H}) are tracers of pre-main-sequence stars \citep{2001ARA&A..39..403R}. They point to the presence of young objects while these are still deeply embedded. Recent observations of brown dwarfs (BDs) yielded evidence for accretion and the presence of circumstellar disks (\cite{2004A&A...424..603N}, \cite{2005Sci...310..834A}, \cite{2007prpl.conf..443L}). Thus they will drive jets and HH flows as well \citep{2004ApJ...615..850M}. Searches for HHOs from very low-mass stars and BDs are being undertaken (e.g. \cite{2005A&A...440.1119F}, \cite{2006ApJ...643..985W}), and the detection of jets from BDs were reported (\cite{2005Natur.435..652W}, \cite{2007astro.ph..3112W}). While these objects are relatively evolved (Class II), our unbiased search for HHOs associated with dark clouds and globules aims at revealing sources which are even younger (Class I). In the following we present three examples of young low-mass objects, for which our investigation yielded new insights and led to a revision of the current knowledge.

\section{Target Selection and Observational Technique}
\label{sec:1}
{The results presented here result from a general search for candidate HHOs using archival data and observations performed with the 2-m telescope of the Th\"uringer Landessternwarte Tautenburg (TLS)}. The archival work is based on DSS-II images and utilises the fact that the $R$ filter almost peaks at H$\alpha$ and the [S{\sc\,ii}]\,$\lambda\lambda$\,6717,\,6731 emission lines. Thus, potential HHOs can be identified in an RGB image based on blue, red, and infrared DSS-II plates since their colours are very different from those of stars. So far the catalogs of \cite{1999ApJS..123..233L} ([LM99]), \cite{2002A&A...383..631D}, and \cite{2005PASJ...57S...1D} were scrutinised for candidates. Since these can be mimicked e.g. by plate artifacts and minor planets, follow-up confirmation is necessary. For this purpose, northern candidates are checked against the $R$ plate of DSS-I which is deep enough for the brighter ones while for southern objects, the SuperCOSMOS H$\alpha$ survey \citep{2005MNRAS.362..689P} is being used. {The TLS-CCD observations aim at the verification of suspected northern HHOs} and a more detailed study. The 2k$\times$2k prime focus CCD camera is used in the Schmidt configuration (diameter of the corrector plate 1.34\,m) for H$\alpha$ and [SII] imaging. It provides a field of view (FOV) of 42\arcmin$\times$42\arcmin{} at the pixel scale of 1\farcs235. Long-slit spectroscopy of the candidate HHOs is obtained using the Nasmyth spectrograph which is equipped with a 2800$\times$800 pixel SITe CCD. A slit width of 1\arcsec{} is normally used which, together with the V100 grism, leads to a resolution of $\rm R\approx2100$. Radial velocities and excitation conditions are derived from the spectra (see \cite{2004ApJ...617..418S} for a more detailed description). 

\subsection{GF9-2 -- The Pre-stellar Core revisited}
\label{sec:3}
\begin{figure}[t]
\begin{center}
\includegraphics[width=0.475\textwidth]{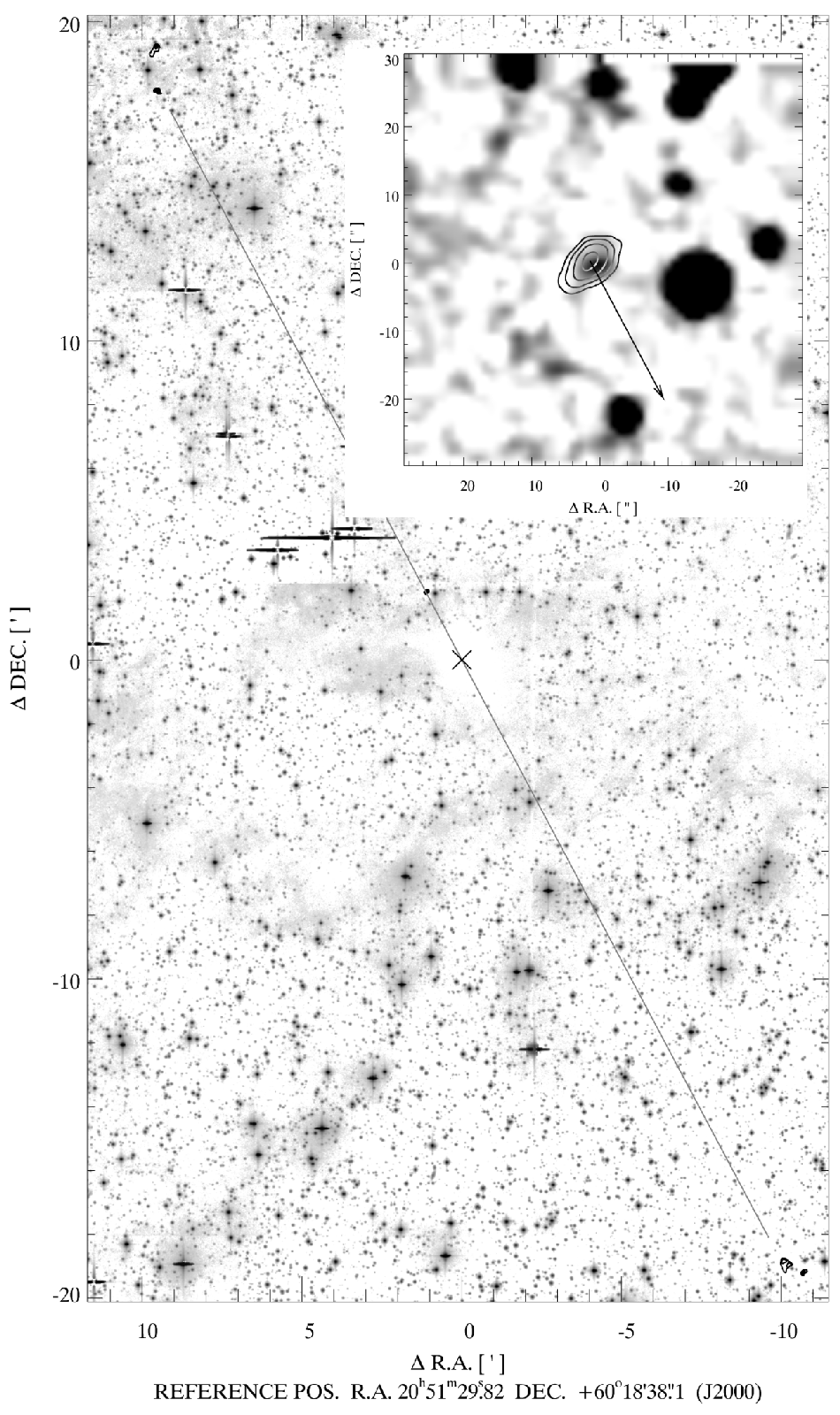}
\caption{{HH flow from GF9-2. The CCD-R-band image is shown together with contours of the continuum-subtracted H$\alpha$ emission. The position of the driving source embedded in GF9-2 is indicated by the inclined cross which corresponds to the location of the 3\,mm source of \cite{2005PASJ...57S...1D}. The grey line marks the flow axis, with the outermost HHOs next to its end points.} The insert shows the DSS-II $R$ image together with continuum-subtracted CCD-H$\alpha$ contours (levels $[3,6,12,24]\sigma$) of the innermost HHO. The arrow points back to GF9-2. The proper motion within $\approx$12\,yrs is evident and directed away from the source.}
\label{fig:1}       % Give a unique label
\end{center}
\end{figure}
GF9-2 is a millimetre source \citep{1994Ap&SS.212..197M} in the globular filament \#9 (\cite{1979ApJS...41...87S}; LDN 1082) associated with IRAS20503+6006. It was classified by \cite{1999usis.conf..533W} as an extremely young source, probably in transition from Class $-$I to 0 {(according to the classification scheme for young stellar objects of \cite{1987ApJ...312..788A} and its extension to Class 0 by \cite{1993ApJ...406..122A}). If true, GF9-2 would represent a rare case of an object turning from the pre-stellar to the proto-stellar phase.} Also \cite{2006ApJ...653.1369F} considered it to be in a very early stage of low-mass star formation because of the non-detection of a molecular outflow. 
%Its bolometric luminosity and temperature are 0.3L$_{\odot}$ and 20\,K, respectively. 
Our search led to the discovery of 14 HHOs in the GF9 region which seem to belong to at least three HH flows. Five HHOs and the millimetre source GF9-2 are very well linearly aligned (correlation coefficient 0.9994, Fig.\,\ref{fig:1}). Thus we conclude that they constitute a HH flow driven by IRAS20503+6006. Its overall length amounts to 43\farcm5 which corresponds to 2.5\,pc for an assumed distance of 200\,pc. {The outer HHOs have almost identical distances from the central source, suggesting a common origin from a past ejection event}. Both south-western components have the morphology of a reversed shock. Remarkably, the position angle of the flow of 28\degr{} is close to that of the projected B field ($\approx$20\degr) derived from ISO polarisation measurements \citep{1999ipo..work....7C}, suggesting a collapse along the field lines. The radial velocity of $v_{\rm LSR}=-69\pm5\rm\,km\,s^{-1}$ of the innermost HHO (Fig.\,\ref{fig:1} insert) indicates that the north-eastern part of the flow is preceding. {If the symmetry of the ejection also holds for this HHO, its distance from the driving source indicates that the receding counterpart is still hidden behind the dark cloud.}  The flow inclination derived from the radial velocity and the proper motion estimate amounts to $\approx$60\degr. The presence of a well developed, {parsec-scale} outflow from GF9-2, which escaped detection in molecular lines so far, indicates a more advanced evolutionary stage of this source than previously believed.

\subsection{L1415-IRS - The least luminous FUor/EXor}
\label{sec:4}
\begin{figure}[t]
\begin{center}
\includegraphics[width=0.4\textwidth]{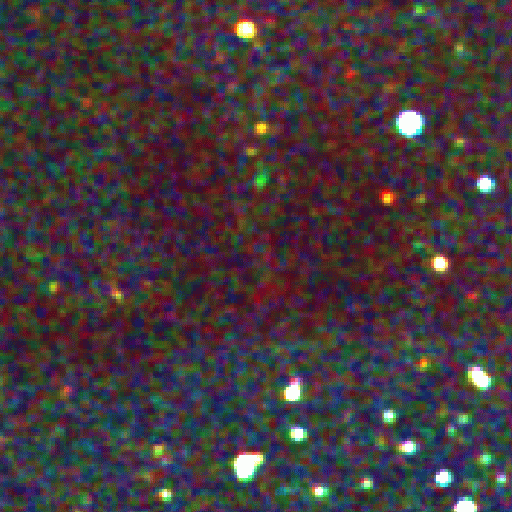}
\includegraphics[width=0.4\textwidth]{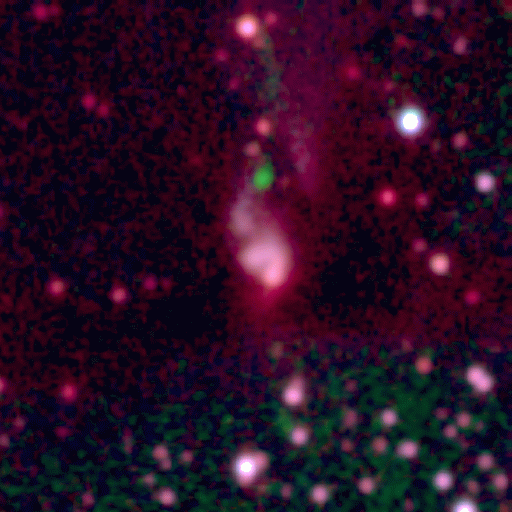}
\caption{Top: DSS-II RGB image of a sub-region of LDN1415 (FOV 1\farcm5$\times$1\farcm5). HH892 appears as green spot. Bottom: I, H$\alpha$, and R colour composite {based on TLS-CCD images.}}
\label{fig:2}       % Give a unique label
\end{center}
\end{figure}
Our CCD observations confirmed the HH nature of a candidate object in LDN1415 (Fig.\,\ref{fig:2}) which is now known as HH 892 \citep{2007A&A...463..621S}. Moreover, a new arcuate nebula associated with IRAS04376+5413 was found {with an integrated $I$-band magnitude of $15.3\pm0.1$\,mag}. The comparison with archival data revealed that it {brightened in $I$ by 3.8\,mag} in recent years. The 2MASS images show a red non-stellar counterpart of the IRAS source, designated as L1415-IRS. The optical spectrum of the nebula displays a pronounced P-Cygni profile across the H$\alpha$ line. These findings represent clear evidence for an FUor- or EXor-type outburst of the embedded object due to temporarily enhanced accretion. The brightening of the nebula results from the enhanced accretion luminosity of the young source and the diminished optical depth due to dust blow-out by the strong neutral wind. The luminosity of L1415-IRS during the inactive state integrated from 0.9\,$\mu$m to 60\,$\mu$m amounts to 0.13\,$L_{\sun}$ for the assumed distance of 170\,pc. It is comparable to that of the low-luminosity source L1014-IRS \citep{2004ApJS..154..396Y}, a possible substellar young object. L1415-IRS is by far the least luminous member of the sparse sample of FUors and EXors. Our monitoring shows that it has been in the active state for {more than} one year now with very minor changes in brightness. Thus, it seems likely that it is of FUor-type rather than being an EXor, {as suggested in the more detailed paper by \citep{2007A&A...463..621S}}. This finding challenges models of the FUor phenomenon based on opacity changes in the disk because of the low effective temperature of the central source. An alternative explanation is accretion bursts due to the capture of protoplanetary bodies was put forward recently by \cite{2006ApJ...650..956V}.

\subsection{BHR111 --  A starless Core remains starless?}
\label{sec:5}

\begin{figure}[ht]
\begin{center}
\includegraphics[width=0.475\textwidth]{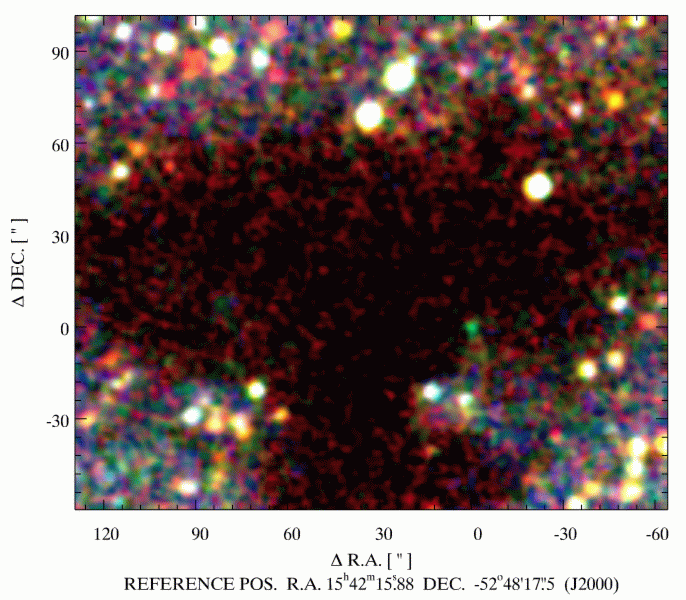}
\caption{DSS-II RGB image of BHR111. The HHO is located at the reference position.}
\label{fig:3}       % Give a unique label
\end{center}
\end{figure}

\begin{figure}[h]
\begin{center}
\includegraphics[width=0.475\textwidth]{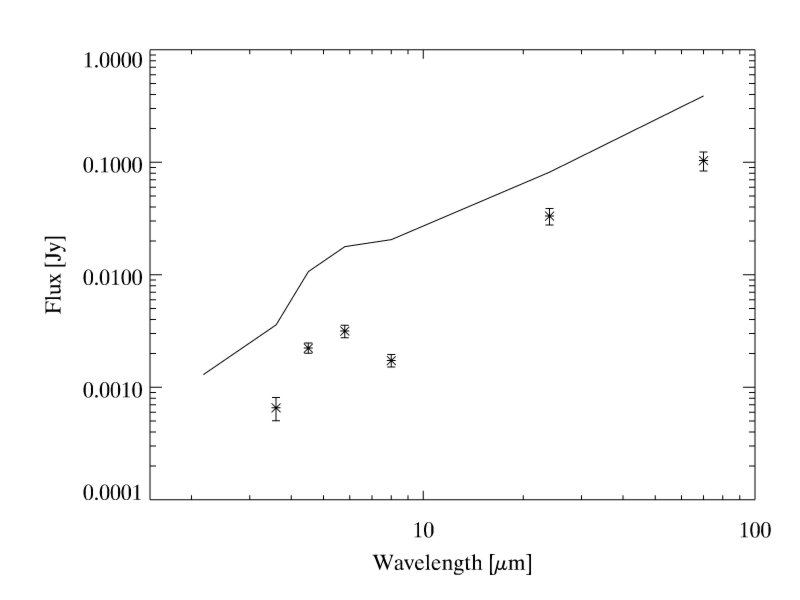}
\caption{SED of the candidate proto-brown dwarf (Stecklum et al, in prep.). The line displays the SED of L1014-IRS \citep{2004ApJS..154..396Y} for comparison.}
\label{fig:4}       % Give a unique label
\end{center}
\end{figure}

Within our {DSS-II} survey a very compact candidate HHO was spotted against the starless core [LM99]135, also known as BHR111 (Fig.\,\ref{fig:3}). The object was detected on the corresponding SuperCosmos H$\alpha$ image which proves its reality. BHR111 is a rather isolated dark cloud. Two IRAS point sources within 7\farcm5 radius are planetary nebula. Thus, we conclude that the HHO points to the presence of a young object within the cloud core which drives an outflow. This is the first case for the detection of an embedded source in a starless core by the identification of an associated HHO. The high stellar density of the surrounding field and the almost complete lack of foreground stars provide compelling evidence that the cloud is nearby. A value of 200\,pc is adopted in the following. The cloud was observed with the {\em Spitzer} infrared space telescope, and the corresponding IRAC and MIPS data were retrieved from the archive. The MIPS 24\,\micron{} image shows three sources next to the core but only the one closest to the HHO is clearly detected at 70\,\micron. Thus we conclude that this object, BHR111-IR, is {likely} driving the outflow. {The detection of a bipolar reflection nebula associated with the {\em Spitzer} source by deep high-resolution near-infrared imaging would rule out that it is a background object}. The photometry derived from the {\em Spitzer} images yielded a spectral energy distribution (Fig.\,\ref{fig:4}) which resembles that of L1014-IRS , suggesting a similar evolutionary phase. {However, the infrared luminosity of 0.01\,L$_{\odot}$ of BHR111-IR is much weaker compared to L1014-IRS (0.09\,L$_{\odot}$) which points to a less massive central source}. 

While is seems probable that L1014-IRS and L1415-IRS will eventually become very low-mass stars since their accretion still continues, BHR111-IRS may stay below the borderline separating BDs from stars indeed despite its ongoing mass assembly. If so the starless core would remain starless (at least for some time).

\section{Outflows from Protoplanets}
The increasing body of evidence for jet and outflow activity of young very-low mass stars and BDs shows that at least a substantial fraction of brown dwarfs form via disk accretion. This raises the question on the minimum mass of an young accreting object that will drive an outflow. The lower mass limit is probably governed by the condition to generate a magnetic field for the acceleration and collimation. Theoretical considerations suggest that a proto-Jovian body in a circumstellar disk might drive an outflow as well (\cite{1998ApJ...508..707Q}, \cite{2003A&A...411..623F}). However,  proto-planetary flows will differ from those of the parent stars because of both the lower mass-loss rate and the orbital motion. The model of \cite{2003A&A...411..623F} implies an H$\alpha$ luminosity of $L_{{\rm H}\alpha}\approx 5\times 10^{-6}\,L_{\odot}$ for an outflow emerging into a neutral environment ($L_{{\rm H}\alpha}$ will be higher by an order of magnitude for a photoionised jet). However, \cite{2003A&A...411..623F} notes that a stellar outflow which expands at its base will disrupt the proto-planetary flow. Therefore, proto-planetary jets can only be observed when the stellar accretion has ceased and planet formation is still ongoing. There might be a time window of a few million years after the clearance of the inner disk when proto-planetary jets can be observable. Extremely large telescopes are required to search for emission lines arising from the biconical surface caused by orbiting proto-planetary jets.

\begin{acknowledgements}
The Digitized Sky Survey was produced at the Space Telescope Science Institute under U.S. Government grant NAG W-2166. The images of these surveys are based on photographic data obtained using the Oschin Schmidt Telescope on Palomar Mountain and the UK Schmidt Telescope. The plates were processed into the present compressed digital form with the permission of these institutions.  Based on observations performed at the Th\"uringer Landessternwarte Tautenburg. This research has made use of NASA's Astrophysics Data System Bibliographic Services and the SIMBAD database, operated at CDS, Strasbourg, France.
\end{acknowledgements}

%% BibTeX users please use the following style file
%\bibliographystyle{Spr-mp-nameyear}
% Here is a sample bibliography file. You will need to substitute your own if you use BibTeX
%\bibliography{5ssr}

\end{document}